\documentclass[12pt,a4paper]{article}
\usepackage{cmap}
\usepackage[T2A]{fontenc}
\usepackage[cp1251]{inputenc}
\usepackage[english, russian]{babel}
\usepackage{amsfonts,amsmath,amssymb,amsthm,longtable,hhline}
\usepackage{bm}
\usepackage{cite}
\usepackage{color,soul}
\usepackage{color,xcolor}
\usepackage{multicol}
\usepackage{tabularx,caption,multirow}
\usepackage{ulem}
\usepackage{titlesec}

\makeatletter
\renewcommand{\@biblabel}[1]{#1.} 
\makeatother
\makeatletter

\newcommand{\clh}[1]{\colorbox{yellow}{#1}}%
\newcommand{\clhp}[1]{\colorbox{yellow}{\parbox{\textwidth}{#1}}}%
\newlength\aptextwidth
\definecolor{BrickRed}{rgb}{0.588,0.098,0.055}

\def\noblue#1{\ifmmode \text{#1}\else #1\fi}
\def\noclh#1{\ifmmode \text{#1}\else #1\fi}
\def\rem#1{}

\let\ge=\geqslant

\def\ttable#1. #2{\begin{table}[t]\tablehat{#1}{#2}}
\def\mtable#1. #2{\begin{table}[hbtp]\tablehat{#1}{#2}}
\def\ptable#1. #2{\begin{table}[p]\tablehat{#1}{#2}}
\def\tablehat#1#2{\centering\small \vbox{\parindent=0pt
  \leftskip=0pt plus.5\hsize \rightskip=\leftskip \parfillskip=0pt
  ТАБЛИЦА #1\\ #2}\nobreak\medskip\medskip }
\def\texendtable{\end{table}}
\def\notation{\par\ifnum\lastpenalty<25000 \bigbreak \fi
  \noindent\triangle\enspace\ignorespaces}

\def\hline{\noalign{\hrule}}

\let\ds=\displaystyle

\let\bl=\bigl \let\br=\bigr
\let\Bl=\Bigl \let\Br=\Bigr
\let\BL=\biggl \let\BR=\biggr

\def\eqnitemskip{\ifhmode \else \par
  \ifnum\lastpenalty>24999
    \ifnum\lastpenalty=25004 \fi
  \else \medbreak \fi \fi }
\def\eqnitem #1. {\eqnitemskip
  {\setbox0=\hbox{$#1^\circ$.\enspace}%
  \ifdim\wd0>\parindent \box0\ignorespaces \else
  \hbox to\parindent{\unhbox0\hss}\ignorespaces\fi}}
\def\eqnitemnobreak #1. {\noindent
  {\setbox0=\hbox{$#1^\circ$.\enspace}%
  \ifdim\wd0>\parindent \box0\ignorespaces \else
  \hbox to\parindent{\unhbox0\hss}\ignorespaces\fi}}
\newdimen\eqnparindent
\eqnparindent=\parindent
\def\eqnitem #1. {\eqnitemskip\noindent\hskip\eqnparindent $#1^\circ$.\enspace\ignorespaces }
\def\eqnitemnobreak #1. {\noindent\hskip\eqnparindent $#1^\circ$.\enspace\ignorespaces }
\def\simpleitem #1. {\eqnitemskip\noindent\hskip\eqnparindent #1.\enspace\ignorespaces }

\def\eqalignno#1{\displ@y \tabskip\centering
  \halign to\displaywidth{\hfil$\@lign\displaystyle{##}$\tabskip\z@skip
    &$\@lign\displaystyle{{}##}$\hfil\tabskip\centering
    &\llap{$\@lign\eqnofont##$}\tabskip\z@skip\crcr
    #1\crcr}}
\let\eqalignno=\eqalignm
\def\eqcenter#1{\displ@y \tabskip\centering
  \halign{\hfil$\displaystyle{##}$\hfil\crcr
    #1\crcr}}
\def\eqcenterno#1{\displ@y \tabskip\centering
  \halign to\displaywidth{\hfil$\@lign\displaystyle{##}$\hfil
    \tabskip\centering&\llap{$\@lign\eqnofont##$}\tabskip\z@skip\crcr
    #1\crcr}}
\def\texcases#1{\left\{\,\vcenter{\normalbaselines\m@th
    \ialign{$##\hfil$&\quad##\hfil\crcr#1\crcr}}\right.}
\def\Displaylines#1{\vcenter{\displ@y \tabskip\z@skip
  \halign{\hbox to\displaywidth{$\@lign\hfil\displaystyle##\hfil$}\crcr
    #1\crcr}}}

\def\tan{\mathop{\operator@font tg}\nolimits}

\def\arb{is an arbitrary constant}
\def\arbs{are arbitrary constants}
\def\arbf{is an arbitrary function}
\def\arbfs{are arbitrary functions}

\makeatother

\titlespacing\section{0pt}{12pt plus 4pt minus 2pt}{0pt plus 2pt minus 2pt}
\titlespacing\subsection{0pt}{12pt plus 4pt minus 2pt}{0pt plus 2pt minus 2pt}

\begin{document}
\large 

\bigskip
\centerline{\bf\Large Closed-form solutions of the nonlinear Schr\"odinger equation}
\centerline{\bf\Large with arbitrary dispersion and potential\clh{$^{*}$}}

\bigskip

\centerline{Andrei D. Polyanin$^*$, Nikolay A. Kudryashov$^{**}$}
 \medskip
\centerline{$^*$~Ishlinsky Institute for Problems in Mechanics, Russian Academy of Sciences,}
\centerline{pr. Vernadskogo 101, bldg. 1, Moscow, 119526 Russia}
\smallskip
\centerline{$^{**}$~Department of Applied Mathematics, National Research Nuclear}
\centerline{University MEPhI, 31 Kashirskoe Shosse,  115409 Moscow, Russia}
\smallskip
\centerline{e-mails: polyanin@ipmnet.ru, nakudryashov@mephi.ru}
\bigskip

\let\thefootnote\relax\footnotetext{
\hskip-20pt\clhp{$^*$ This is a preprint of an article that will be published in the journal \textbf{Chaos, Solitons \& Fractals, 2025, Vol.~191, 115822};
doi:\! 10.1016/j.chaos.2024.115822.}}

\vspace{5.0ex}

For the first time, the general nonlinear Schr\"odinger equation is investigated, in which the chromatic dispersion and potential are specified by two arbitrary functions.
The equation in question is a natural generalization of a wide class of related nonlinear partial differential equations that are often used in various areas of theoretical physics, including nonlinear optics, superconductivity and plasma physics.
To construct exact solutions, a combination of the method of functional constraints and methods of generalized separation of variables is used.
Exact closed-form solutions of the general nonlinear Schr\"odinger equation, which are expressed in quadratures or elementary functions, are found. One-dimensional non-symmetry reductions are described, which lead the considered nonlinear partial differential equation to a simpler ordinary differential equation or a system of such equations.
The exact solutions obtained in this work can be used as test problems intended to assess the accuracy of numerical and approximate analytical methods for integrating nonlinear equations of mathematical physics.
\medskip

\textsl{Keywords\/}:
nonlinear Schr\"odinger equations,
exact closed-form solutions,
solutions in quadratures,
solution in elementary functions,
method of functional constraints,
non-symmety reductions

\section{Introduction}

\subsection{Nonlinear Schr\"odinger equations}

$1^\circ$.
In many areas of theoretical physics, one encounters the classical nonlinear Schr\"odinger equation with cubic nonlinearity
(see, for example, \cite{Ag1, Ag2, Ag3, Ag4, Ag5}):
\begin{equation}
iu_t+u_{xx}+k|u|^2u=0,
\label{Schrodinger-eq}
\end{equation}
where $u=u(x,t)$ is the desired complex-valued function of real variables,
the square of the module of which determines the intensity of light,
$t$ is the time, $x$ is the spatial variable, $k$ is a parameter, $i^2=-1$.

Eq. \eqref{Schrodinger-eq} is used for mathematical modeling of wave propagation in almost all areas of physics where nonlinear wave processes are considered.
The theoretical and experimental justification for the use of the nonlinear Schr\"odinger equation in non-linear optics is given in \cite{UFN, Has1, Has2}. When describing the propagation of pulses in an optical fiber, the expression with the second derivative is responsible for the dispersion of the pulse, the quadratic function $g(|u|)=k|u|^2$, which sets the potential, is called the Kerr nonlinearity.
It characterizes the interaction of a light pulse with a fiber material and specifies the law of refraction of light in a nonlinear medium.
The uniqueness of Eq. \eqref{Schrodinger-eq} is explained not only by the fact that this partial differential equation (PDE) is the basic equation for describing the processes of information transfer in an optical medium, but also by the fact that it belongs to the class of integrable partial differential equations \cite{Ag5}. Eq. \eqref{Schrodinger-eq} has an infinite number of conservation laws, B\"acklund transformations, and passes the Painlev\'e test for partial differential equations \cite{Ag4, Ag5, Pain_3, Pain_4, Pain_5}. The Cauchy problem for Eq.~\eqref{Schrodinger-eq} with an initial condition of the general form is solved by the inverse scattering transform \cite{Ag4, Ag5}. The exact solutions of the nonlinear Schr\"odinger equation \eqref{Schrodinger-eq} are given in \cite{polzai2012, kha2019, pol2025}.

$2^\circ$.
Related nonlinear partial differential equations of the form
\begin{equation}
iu_t+u_{xx}+g(|u|)u=0,
\label{Schrodinger-eq2}
\end{equation}
and other nonlinear Schr\"odinger-type equations that occur in the scientific literature can be found, for example, in
\cite{polzai2012, kha2019, pol2025, Kudr_1, Kudr_2, Kudr_3, Kudr_4, Bis_1, Bis_2, Bis_3, dan2021, Bis_5, Kud2022,mal2022, Bis_6,wan2023,wan2023a,li2024,zho2024,kud2023x,arn2023a,arn2023b,eki2022,zay2024,han2023}
(see also Section 2 of this article).
In nonlinear optics, the potential $g(|u|)$ in Eq. \eqref{Schrodinger-eq2} determines
the law of interaction of a light pulse with a fiber material. Exact solutions of Eq. \eqref{Schrodinger-eq2} in the case of power dependence
$g(|u|)=k|u|^n$ have been considered, for example, in \cite{polzai2012,kha2019,pol2025}.
In plasma theory and laser physics, the PDE \eqref{Schrodinger-eq2} with $g(|u|)=k(1-e^{a|u|})$ is encountered (see, for example, \cite{bol1978}).
Exact closed-form solutions of the nonlinear Schr\"odinger equation \eqref{Schrodinger-eq2} for arbitrary function $g(|u|)$ are given in \cite{polzai2012,pol2025}.

In this article, we will consider a nonlinear PDE that is much more complex than equation \eqref{Schrodinger-eq2}. In this equation, instead of the linear dispersion term given by the second derivative $u_{xx}$, we will include a nonlinear dispersion of the form $[f(u)u_x]_x$, where $f(u)$ is an arbitrary function.
Thus, two arbitrary functions $f(u)$ and $g(u)$ will enter into the nonlinear PDE under consideration at once.

\subsection{Exact solutions of nonlinear partial differential equations\\ (terminology)}

In this article, exact solutions of PDEs are understood as the following solutions \cite{polzai2012,polzhu2022}:

(\textit{a})\enspace Solutions which are expressed via the elementary functions.

(\textit{b})\enspace Solutions which are expressed in quadratures, i.e. through elementary functions, functions included in the equation (this is necessary if the equation contains arbitrary or special functions), and indefinite integrals.

(\textit{c})\enspace Solutions that are expressed through solutions of ordinary differential equations (ODEs) or systems of such equations.

Various combinations of solutions described in Items  (\textit{a})--(\textit{c}) are also allowed. In cases (\textit{a}) and (\textit{b}), the exact solution can be presented in explicit, implicit, or parametric form.

Note that exact solutions play
an important role as mathematical standards which are often used as test problems to check the adequacy and assess the accuracy of numerical and approximate analytical methods for integrating nonlinear PDEs. The most preferable for these purposes are simple solutions from Items (\textit{a}) and (\textit{b}).
Several such exact solutions are described later in this article.
\medskip

\textit{Remark 1.}
Of great interest are also solutions of nonlinear PDEs, which are expressed through solutions of linear PDEs
(examples of such nonlinear PDEs can be found, for example, in \cite{polzai2012,ibr1994,bro2023}),
as well as conditionally integrable nonlinear PDEs \cite{bro2023} (which can also be called partially linearizable PDEs).
\medskip

Exact closed-form solutions of nonlinear PDEs are most often constructed using the classical method of symmetry reductions~(see, for example, \cite{ovs1982,ibr1994,blu1989,olv2000,van2012,polaks2024}),
the direct method of symmetry reductions~(see, for example, \cite{polzai2012,cla1989,nuc1992,olv1994,cla1997,polzhu2022}),
the nonclassical methods of symmetry reductions~(see, for example, \cite{cla1997,blu1969,arr1993,puc2000,bra2019,cher2017}),
methods of generalized separation of variables (see, for example, \cite{polzai2012,galsvi2007,polzhu2022,gal1995,pol2001b,kos2020,polaks2024}),
methods of functional separation of variables (see, for example, \cite{polzai2012,polzhu2022,puc2000,mil1993,doy1998,est2002,pol2020}),
the~method of differential constraints (see, for example, \cite{polzai2012,olv1994,polzhu2022,sid1984,kap2003,mel2005,kru2008}),
the inverse scattering method (see, for example, \cite{Ag5,abl1974,cal1982,kuo2023,vu2023}),
the Painlev\'e analysis of  nonlinear PDEs (see, for example, \cite{jim1982,wei1983a,wei1983b,con1989,kud2010,con2020}),
and some other exact analytical methods~(see, for example, \cite{polzai2012,kud2005,aks2021,polsorzhu2024}).

It is important to note that for complex nonlinear PDEs depending on one or more arbitrary functions (and it is precisely such nonlinear PDEs that are considered in this article), the vast majority of existing analytical methods for constructing exact solutions are either not applicable at all or are weakly effective.
Statistical processing of reference data \cite{polzai2012,pol2025} showed that at present the majority of exact solutions of such equations were obtained by methods of generalized and functional separation of variables. Significantly fewer exact solutions of such equations were obtained by nonclassical methods of symmetry reductions and the method of differential constraints, which are much more difficult to use in practice.
In general, very few nonlinear PDEs depending on arbitrary functions that admit exact closed-form solutions are known today.

\section{General nonlinear Schr\"odinger equation. Transformations\\ to a system of real PDEs}\label{s:2}

\subsection{The nonlinear Schr\"odinger equation with arbitrary dispersion and potential}

$1^\circ$. In this paper we consider the nonlinear Schr\"odinger equation of the general form
\begin{equation}
iu_t+[f(|u|)u]_{xx}+g(|u|)u=0,
\label{eq01}
\end{equation}
where $u=u(x,t)$ is the complex-value function of real variables, $f(z)$ and $g(z)$ are arbitrary real functions ($f$ is a twice continuously differentiable function, and $g$ is a continuous function), $i^2=-1$.
Eq.~\eqref{eq01} with nonlinear dispersion, which is given by the function $f(z)$, is an essential generalization of Eq.~\eqref{Schrodinger-eq2} with
$f(z)=\text{const}$.

The nonlinear Schr\"odinger equation \eqref{eq01} with power functions
\begin{equation}
f(z)=az^k,\quad \ g(z)=bz^m+cz^n
\label{eq01*}
\end{equation}
has been considered in \cite{Kud2022}, where its solutions
of the type of stationary solitons of the form $u=r(x)e^{i\lambda t}$ were described.
Some exact solutions of Eq.~\eqref{eq01} in the case of arbitrary functions $f(|u|)$ and $g(|u|)\equiv 0$
are presented in \cite{polzai2012}.

$2^\circ$. Note that if $u(x,t)$ is a solution of equation \eqref{eq01}, than the functions
\begin{equation}
\bar u=\pm e^{iC_1}u(\pm x+C_2,t+C_3),
\label{eq01**}
\end{equation}
where $C_1$, $C_2$, $C_3$ are arbitrary real constants,
are also solutions of this equation  (the signs in Eq.~\eqref{eq01**} can be chosen independently of each other).
From this property follows, that Eq. \eqref{eq01} admits traveling wave solutions of the form
$u=U(z)$, \text{$z=x-\lambda t$}, where $\lambda$ \arb \ (more complex solutions including traveling wave solutions are given in Section~\ref{s:4}).

\subsection{Transformation of the general nonlinear Schr\"odinger equation to a system of real PDEs}

Let us represent the desired function in exponential form
\begin{equation}
u=re^{i\varphi},\quad \ r=|u|,
\label{eq02}
\end{equation}
where $r=r(x,t)\ge 0$ and $\varphi=\varphi(x,t)$ are real functions.

Differentiating \eqref{eq02}, we find the derivatives:
\begin{equation}
\begin{aligned}
u_t&=(r_t+ir\varphi_t)e^{i\varphi},\\
[f(|u|)u]_{x}&=(h_x+ih\varphi_x)e^{i\varphi},\quad h=rf(r),\\
[f(|u|)u]_{xx}&=[h_{xx}-h\varphi_x^2+i(2h_x\varphi_x+h\varphi_{xx})]e^{i\varphi},\quad h=rf(r).
\end{aligned}
\label{eq03}
\end{equation}
Now we substitute \eqref{eq03} into \eqref{eq01}, and then divide all terms by $e^{i\varphi}$.
Further equating the real and imaginary parts of the obtained relation to zero, we arrive at the following system of two real partial differential equations:
\begin{equation}
\begin{aligned}
-r\varphi_t+h_{xx}-h\varphi_x^2+rg(r)&=0,\\
r_t+2h_x\varphi_x+h\varphi_{xx}&=0,\quad \ h=rf(r).
\end{aligned}
\label{eq04}
\end{equation}

We use system of equations \eqref{eq04} and expressions \eqref{eq02} to construct exact solutions of the nonlinear Schr\"odinger equation \eqref{eq01}.
\medskip

\textit{Remark 2.}
Note that the second equation of system \eqref{eq04} can be written in the form of conservation law
\begin{equation}
H_t+(h^2\,\varphi_x)_x=0
\label{eq04a}
\end{equation}
where the notation is used
\begin{equation*}
H=\int h\,dr.
\end{equation*}

From equation \eqref{eq04a}, under some standard restrictions, the law of conservation of density follows, $\int^\infty_{-\infty}H\,dx=\text{const}$.
In addition, the conservation law \eqref{eq04a} can be used to evaluate the accuracy of the results of applying numerical methods of integrating the system \eqref{eq04}.
\medskip

\subsection{Method of constructing exact solutions based on functional\\ constraints}

The construction of exact closed-form solutions of Eq. \eqref{eq01} is complicated by the fact that it contains two arbitrary
functions $f(z)$ and $g(z)$.
For such equations and other nonlinear PDEs of the general form that contain arbitrary functions, it is impossible to directly use the standard methods of generalized separation of variables described in books \cite{polzai2012,polzhu2022,galsvi2007}.

To find exact solutions of the general nonlinear Schr\"odinger equation \eqref{eq01}, we impose one of the three additional relations (functional constraints) on the argument of arbitrary functions:
\begin{align}
|u|&=\text{const},\label{eq02aa}\\
|u|&=p(x),\label{eq02ab}\\
|u|&=q(t),\label{eq02ac}
\end{align}
where $p(x)$ and $q(t)$ are some function of one argument.
When any of the relations \eqref{eq02aa}--\eqref{eq02ac} is satisfied, equation \eqref{eq01} is ``linearized''.
This important fact allows us to further use
the standard approach for separating variables applied for linear PDEs or methods of generalized separation of variables used for nonlinear PDEs \cite{polzai2012,polzhu2022,galsvi2007}.
A similar technique,
based on the use of additional relations of the type \eqref{eq02aa}--\eqref{eq02ac} and called the method of functional constraints, has made it possible to find many exact solutions of nonlinear PDEs with delay \cite{polzhu2014a,polzhu2014b,polsor2022,polsorzhu2024}.

After the transition from the complex equation \eqref{eq01} to the system of real PDEs \eqref{eq04}, when constructing exact solutions, one should use the relations
\eqref{eq02aa}--\eqref{eq02ac}, setting in them $|u|=r$ (this follows from representation \eqref{eq02}).
Thus, the functions $p(x)$ and $q(t)$ in \eqref{eq02ab} and \eqref{eq02ac} depend on $x$ and $t$ implicitly and are expressed through the amplitude $r$.
\medskip

\textit{Remark 3.}
In Section \ref{s:4} (see Remark~4) it will be shown that in addition to \eqref{eq02aa}--\eqref{eq02ac}, a more complex functional constraint based on the transition to a new variable of the traveling wave type can also be used to construct exact solutions.

\section{Exact solutions of the general nonlinear Schr\"odinger equation}\label{s:3}

In this section exact closed-form solutions of the general nonlinear Schr\"odinger equation \eqref{eq01}
with arbitrary functions $f(z)$ and $g(z)$ are given.
To construct these exact solutions we use the functional constraints \eqref{eq02aa}--\eqref{eq02ac} and
the methods of generalized separation of variables.

\subsection{Traveling wave solutions with constant amplitude}

First, we use the simplest additional relation \eqref{eq02aa}.
Given that $|u|=r$, we substitute $r=C_1=\text{const}$ into system \eqref{eq04}.
From the second equation \eqref{eq04} it follows that $\varphi''_{xx}=0$.
From here, after integration, we have $\varphi(x)=a(t)x+b(t)$, where $a(t)$ and $b(t)$ are arbitrary functions.
Substituting this expression into the first equation \eqref{eq04}, we find $a(t)=C_2$ and $b(t)=Bt+C_3$.
As a result, we arrive at the simple exact solution of the PDE system \eqref{eq04}:
\begin{equation}
r=C_1,\quad \ \varphi=C_2x+C_3+Bt,\quad \ B=g(C_1)-C_2^2f(C_1),
\label{eq05ab}
\end{equation}
where  $C_1$, $C_2$, and $C_3$ are arbitrary real constants ($C_1>0$). Substituting \eqref{eq05ab} into \eqref{eq02}, we have the traveling wave solution
of the nonlinear PDE~\eqref{eq01}:
\begin{equation}
u=C_1e^{i(C_2x+C_3+Bt)},\quad \ B=g(C_1)-C_2^2f(C_1).
\label{eq05abc}
\end{equation}
This is a solution that is periodic in time and space with constant amplitude $C_1$.

In the case of power functions \eqref{eq01*} solution \eqref{eq05abc} takes the form
\begin{equation*}
u=C_1e^{i(C_2x+C_3+Bt)},\quad \ B=bC_1^m+cC_1^n-aC_1^kC_2^2.
\end{equation*}

\subsection{Time-periodic solutions with amplitude depending on the spatial variable $x$}

To construct another exact solution, we use the additional relation \eqref{eq02ab}. Given that $|u|=r$,
we substitute $r=r(x)$ into system \eqref{eq04}. Integrating the second equation of the resulting system twice, we have
\begin{equation}
\varphi=a(t)\int h^{-2}dx+b(t),
\label{eq30}
\end{equation}
where $a=a(t)$ and $b=b(t)$ \arbfs. By eliminating $\varphi$ from the first equation of system \eqref{eq04} using \eqref{eq30},
we arrive at an equation that, after dividing all terms by $r$, takes the form
\begin{equation}
\begin{aligned}
-a'_t\int h^{-2}dx-b'_t-a^2r^{-1}h^{-3}+r^{-1}[h_{xx}''+rg(r)]&=0,\quad \ h=rf(r).
\end{aligned}
\label{eq31}
\end{equation}

To solve the functional differential equation \eqref{eq31}, which includes the desired functions depending on different variables, we use the methods of generalized separation of variables \cite{polzai2012,polzhu2022}. Standard analysis shows that two different situations are possible, described below.

$1^\circ$. \textit{The general case when two functions $f$ and $g$ are arbitrary}.
Setting in \eqref{eq31}
\begin{equation}
a(t)=C_2,\quad \ b(t)=C_1t+C_3,
\label{eq32}
\end{equation}
where $C_1$, $C_2$, and $C_3$ \arbs, we obtain a second-order nonlinear ODE of the autonomous form
\begin{equation}
h_{xx}''-C_2^2h^{-3}-C_1r+rg(r)=0,\quad \ h=rf(r).
\label{eq08}
\end{equation}

Let us first consider the degenerate case when $h''_{xx}\equiv 0$, i.e.
\begin{equation}
h(x)=ax+b,
\label{eq70}
\end{equation}
where $a$ and $b$ are arbitrary constants.
We substitute the function \eqref{eq70} into \eqref{eq08} and take into account the relation $h=rf(r)$.
As a result, we obtain a potential, which in this case is expressed through the function $f(r)$ as
\begin{equation}
g(r)=C_1+C_2^2r^{-4}f^{-3}(r).
\label{eq50}
\end{equation}
It follows that of the nonlinear Schr\"odinger equation
\begin{equation}
iu_t+[f(|u|)u]_{xx}+[C_1+C_2^2|u|^{-4}f^{-3}(|u|)]u=0,
\label{eq51aa}
\end{equation}
where $f(r)$ is an arbitrary function, has an exact solution of the form \eqref{eq02},
in which the amplitude $r=r(x)$ is given by the implicit relation
\begin{equation}
ax+b=rf(r),
\label{eq52aa}
\end{equation}
and the phase is found using formula \eqref{eq07} and can be written in the form
\begin{equation}
\varphi=C_1t+C_2\int h^{-2}dx+C_3=C_1t-\frac{C_2}{a(ax+b)}+C_3.
\label{eq53aa}
\end{equation}

In the general case of two arbitrary functions $f(r)$ and $g(r)$, the general solution of the second-order nonlinear ODE \eqref{eq08} can be expressed in quadratures in implicit form (detailed explanations are given in Appendix):
\begin{equation}
\begin{gathered}
\int\BL[-C_2^2h^{-2}+2\int[C_1r- rg(r)]\,dh+C_4\BR]^{-1/2}\,dh=C_5\pm x,\\
h=rf(r),\quad \ dh=[f(r)+rf'_r(r)]\,dr,
\end{gathered}
\label{eq08**}
\end{equation}
where $C_4$ and $C_5$ \arbs.

In this case, the phase $\varphi$ is expressed through the amplitude $r$ by the formulas
\begin{equation}
\varphi=C_1t+C_2\int h^{-2}dx+C_3=C_1t+C_2\int[rf(r)]^{-2}dx+C_3,
\label{eq07}
\end{equation}
which are obtained by substituting the expressions \eqref{eq32} into \eqref{eq30}.

Thus, we have shown that the PDE system \eqref{eq04} admits a time-periodic solution,
which is expressed in quadratures using the relations \eqref{eq08**} and \eqref{eq07}.

Setting $C_2=0$ in \eqref{eq08**} and \eqref{eq07}, we obtain solutions describing stationary solitons,
whose phase $\varphi$ does not depend on the spatial coordinate $x$.
\medskip

\textit{Example 1.}
Let us consider the nonlinear Schr\"odinger equation \eqref{eq01} with a constant potential,
\begin{equation}
g(r)=C_1=\text{const}.
\label{eq80}
\end{equation}
Substituting \eqref{eq80} into solution \eqref{eq08**}, after integration we have
\begin{equation}
\sqrt{C_4h^2-C_2^2}=C_4(C_5\pm x).
\label{eq81}
\end{equation}
By squaring both parts \eqref{eq80} and taking into account the formula $h=rf(r)$, we obtain the relation
\begin{equation}
C_4r^2f^2(r)=C_2^2+C_4^2(C_5\pm x)^2,
\label{eq82}
\end{equation}
which implicitly determines the dependence of the amplitude $r$ on $x$.
Using formulas \eqref{eq07} and \eqref{eq81}, we find the phase
\begin{equation}
\varphi=C_1t+\arctan\BL[\frac{C_4(x\pm C_5)}{C_2}\BR]+C_3.
\label{eq07xx}
\end{equation}

Note that in equation \eqref{eq01} (with \eqref{eq80}) and solution \eqref{eq82} the function $f=f(r)$ can be specified arbitrarily.
\medskip


The general solution of equation \eqref{eq08}, represented in quadratures in the implicit form \eqref{eq08**},
is difficult to analyze due to the presence of two arbitrary functions in it $f(r)$ and $g(r)$.
Therefore, we will describe here a simpler and more convenient for analysis inverse (not direct) approach, based on introducing another arbitrary auxiliary function instead of the function $g(r)$ and directly specification of exact solutions in implicit form.
Namely, we will consider the function $f(r)$ to be arbitrary, and the solution $r=r(x)$ of equation \eqref{eq08}, taking into account the relation $h=rf$, will be specified using an arbitrary function $h=h(x)$ in implicit form
\begin{equation}
rf(r)=h(x).
\label{eq90}
\end{equation}
The function $g=g(r)$ in this approach is found from equation \eqref{eq08}, that leads to the formula
\begin{equation}
g=C_1+r^{-1}[C_2^2h^{-3}(x)-h''_{xx}(x)].
\label{eq91}
\end{equation}
The potential function $g=g(r)$ for a given specific function $h(x)$ is determined by eliminating $x$ from relations \eqref{eq90}--\eqref{eq91}.

Let us demonstrate with several specific examples how the described approach works in practice.
To do this, we will take simple elementary functions $h=h(x)$ and will find the potential functions $g=g(r)$ generated by them.
In these cases, the functions $g(r)$ will be expressed through $f(r)$.
\medskip

\textit{Example 2.}
In relations \eqref{eq90}--\eqref{eq91} we substitute the function
\begin{equation}
h(x)=a(x+b)^k,
\label{eq92}
\end{equation}
where $a$, $b$, and $k$ \arbs.
As a result, we have
\begin{equation}
\begin{aligned}
&rf(r)=a(x+b)^k,\\
&g=C_1+r^{-1}[a^{-3}C_2^2(x+b)^{-3k}-ak(k-1)(x+b)^{k-2}].
\end{aligned}
\label{eq93}
\end{equation}
By excluding $x$ from these relations, we obtain the dependence of the potential on~$r$:
\begin{equation}
g(r)=C_1+C_2^2r^{-4}f^{-3}(r)-a^{2/k}k(k-1)r^{-2/k}f^{(k-2)/k}(r).
\label{eq94}
\end{equation}

In particular, by setting $C_2=0$ and $k=-1$ in \eqref{eq94}, we arrive at the nonlinear Schr\"odinger equation
\begin{equation}
iu_t+[f(|u|)u]_{xx}+[C_1-2a^{-2}r^2f^3(|u|)]u=0.
\label{eq95}
\end{equation}
This equation, containing an arbitrary function $f(|u|)$, admits an exact solution of the form
\begin{equation}
u=r(x)e^{i(C_1t+C_3)},
\label{eq52}
\end{equation}
in which the amplitude $r=r(x)$ is given by the implicit relation
$$
rf(r)=a(x+b)^{-1}.
$$

\textit{Example 3.}
In relations \eqref{eq90}--\eqref{eq91} we substitute $h=a\cos(kx+b)$ and \text{$C_2=0$}.
Taking into account that $h''_{xx}=-k^2h$ and $h=rf(r)$, we find the potential $g=C_1+k^2f(r)$.
Thus, we arrive at the nonlinear Schr\"odinger equation
\begin{equation}
iu_t+[f(|u|)u]_{xx}+[C_1+k^2f(|u|)]u=0,
\label{eq51}
\end{equation}
which has an exact solution of the form \eqref{eq52}, where the amplitude $r=r(x)$ is given by the implicit expression
$$
a\cos(kx+b)=rf(r).
$$

\textit{Example 4.}
Setting $h=ae^{kx}+be^{-kx}$ and $C_2=0$ in \eqref{eq90}--\eqref{eq91}, we obtain the Schr\"odinger equation
\begin{equation}
iu_t+[f(|u|)u]_{xx}+[C_1-k^2f(|u|)]u=0,
\label{eq53}
\end{equation}
which has an exact solution of the form \eqref{eq52}, where the amplitude $r=r(x)$ is given implicitly by the relation
$$
ae^{kx}+be^{-kx}=rf(r).
$$

$2^\circ$. \textit{The case when two functions $f$ and $g$ are given by one arbitrary function}.
Equation \eqref{eq31} includes two time-dependent functions $a=a(t)$ and $b=b(t)$.
Using the splitting principle \cite{polzai2012,polzhu2022}, we impose the following differential constraints on these functions:
\begin{equation}
a'_t=-A_1a^2+B_1,\quad \ b'_t=-A_2a^2+B_2,
\label{eq33}
\end{equation}
where $A_1$, $B_1$, $A_2$, and $B_2$ \arbs.
As a result, we have
$$
a^2\BL(A_1\!\int h^{-2}dx+A_2-r^{-1}h^{-3}\BR)-B_1\!\int h^{-2}dx-B_2+r^{-1}[h_{xx}''+rg(r)]=0.
$$
By equating the functional factor at $a^2$ to zero, we obtain two integro-differential equations
\begin{align}
A_1&\!\int h^{-2}dx+A_2-r^{-1}h^{-3}=0,\quad \ h=rf(r),\label{eq34}\\
-B_1&\!\int h^{-2}dx-B_2+r^{-1}[h_{xx}''+rg(r)]=0.\label{eq35}
\end{align}
Depending on the value of the free coefficient
$A_1$, two different situations may arise.
Let us consider them in order.

2.1. For $A_1\not=0$,
eliminating the integral term from \eqref{eq34}--\eqref{eq35}, we arrive at the ODE:
\begin{equation}
h_{xx}''-\frac{B_1}{A_1}h^{-3}+\Bl(B_1\frac{A_2}{A_1}-B_2\Br)r+rg(r)=0,\quad \ h=rf(r),
\label{eq36}
\end{equation}
which, up to renaming of constants, coincides with equation \eqref{eq08}.
Instead of equation \eqref{eq35}, we can consider equation \eqref{eq36}.
Note that equation \eqref{eq34} reduces to an ODE by differentiating with respect to $x$.

The system of two equations \eqref{eq34}--\eqref{eq35} includes three functions $r=r(x)$, $f=f(r)$, and $g=g(r)$.
If we set one of these functions arbitrarily, then the other two can be found from equations \eqref{eq34}--\eqref{eq35}
or equations \eqref{eq34} and \eqref{eq36}.

The solution of system \eqref{eq34}--\eqref{eq35} can be represented in parametric form by arbitrarily setting the function $h=h(x)$ (which is the product of the functions $r$ and $f$). Indeed, substituting this function into equations \eqref{eq34} and \eqref{eq36}, we obtain
\begin{equation}
\begin{aligned}
r&=h^{-3}(x)\BL[A_1\int\frac{dx}{h^2(x)}+A_2\BR]^{\!-1},\\
f&=h^4(x)\BL[A_1\int\frac{dx}{h^2(x)}+A_2\BR],\\
g&=B_2-B_1\frac{A_2}{A_1}+\Bl[\frac{B_1}{A_1}-h^3(x)h_{xx}''(x)\Br]\BL[A_1\int\frac{dx}{h^2(x)}+A_2\BR],
\end{aligned}
\label{eq37}
\end{equation}
where $h=h(x)$ \arbf, and $x$ plays the role of a parameter here.

For $A_1\not=0$, the functions $a=a(t)$ and $b=b(t)$, included in the formula for phase \eqref{eq30},
are determined by integrating the first-order ODEs \eqref{eq33} and are expressed through elementary functions:
\begin{align}
a(t)&=\begin{cases}\ds\frac{B_1}{\sqrt{A_1B_1}}\tanh\bl(\sqrt{A_1B_1}\,t+C_1\br)& \text{if} \ A_1B_1>0,\\
\ds\frac{B_1}{\sqrt{|A_1B_1|}}\tan\bl(\sqrt{|A_1B_1|}\,t+C_1\br)& \text{if} \ A_1B_1<0,\\
(A_1t+C_1)^{-1}& \text{if} \ B_1=0;\end{cases}\label{eq38}\\
b(t)&=\begin{cases}
\ds\BL(B_2-A_2\frac{B_1}{A_1}\BR)t+\frac{A_2B_1}{A_1\sqrt{A_1B_1}}\tanh\bl(\sqrt{A_1B_1}\,t+C_1\br)+C_2& \text{if} \ A_1B_1>0,\\
\ds\BL(B_2-A_2\frac{B_1}{A_1}\BR)t+\frac{A_2B_1}{A_1\sqrt{|A_1B_1|}}\tan\bl(\sqrt{|A_1B_1|}\,t+C_1\br)+C_2& \text{if} \ A_1B_1<0,\\
\ds\frac{A_2}{A_1(A_1t+C_1)}+B_2t+C_2& \text{if} \ B_1=0,\end{cases}\label{eq39}
\end{align}
where $C_1$ and $C_2$ \arbs.
\medskip

\textit{Example 5.}
Substitute
$$
A_2=0,\quad \ h=k(x+C_3)^{-1/2},
$$
into \eqref{eq37}. After elementary calculations we get
\begin{align*}
r&=\frac 2{A_1k}(x+C_3)^{-1/2},\quad \ f=\frac 12 A_1k^2=\text{const},\\
g&=B_2+\frac 12 B_1k^{-2}(x+C_3)^2-\frac 38A_1k^2(x+C_3)^{-2}.
\end{align*}
Eliminating $x$ from here, and for simplicity setting $k=\sqrt{2/A_1}$, we arrive at Eq.~\eqref{Schrodinger-eq2} with
\begin{equation}
g(r)=B_2+\frac{B_1}{A_1}r^{-4}-\frac 3{16}A_1^2r^4,\quad \ r=|u|.
\label{eq39*}
\end{equation}
Equation \eqref{Schrodinger-eq2} with potential \eqref{eq39*} has the exact solution
$$
u=re^{i\varphi},\quad \ r=\sqrt{\frac 2{A_1(x+C_3)}},\quad \ \varphi=\frac14 A_1a(t)(x+C_3)^2+b(t),
$$
where the functions $a=a(t)$ and $b(t)$ are determined by formulas \eqref{eq38} and \eqref{eq39}.
\medskip

2.2. When $A_1=0$, it follows from equation \eqref{eq34} that $f(r)=A_2^{-1/3}r^{-4/3}$.
Equation \eqref{eq35} can be reduced by differentiation to a third-order nonlinear ODE, which cannot be integrated in closed form.
Therefore, to construct some exact solutions of the nonlinear Schr\"odinger equation \eqref{eq01} we use here
the inverse approach (without integrating ODEs) described earlier.
For this we first set an arbitrary function $h=h(x)$.
Substituting this function into equations \eqref{eq34}--\eqref{eq35} with $A_1=0$, we obtain
\begin{equation}
\begin{aligned}
r&=A_2^{-1}h^{-3}(x),\\
f&=A_2h^4(x),\\
g&=B_1\int h^{-2}(x)\,dx+B_2-A_2h^3(x)h''_{xx}(x).
\end{aligned}
\label{eq40**}
\end{equation}
These formulas define the solution of system \eqref{eq34}--\eqref{eq35} with $A_1=0$, which is given in parametric form using one arbitrary function $h=h(x)$. The dependence of the potential $g=g(r)$ is obtained by eliminating $x$ from the first and third relations \eqref{eq40**}.

For $A_1=0$, the functions $a=a(t)$ and $b=b(t)$, included in the formula for phase \eqref{eq30}, are determined by integrating the first-order ODEs \eqref{eq33} and are expressed through elementary functions:
$$
a(t)=B_1t+C_1,\quad \ b(t)=-\frac {A_2}{3B_1}(B_1t+C_1)^3+B_2t+C_2,
$$
where $C_1$ and $C_2$ \arbs.

\subsection{Generalized separable solutions with amplitude depending on the time $t$}

To construct an exact solution, we now use the additional relation \eqref{eq02ac}. Given that $|u|=r$, we substitute $r=r(t)$ into the PDE system \eqref{eq04}. By integrating the second equation of the resulting system twice, we arrive at a solution with generalized separation of variables of the following form:
\begin{equation}
r=r(t),\quad \ \varphi=a(t)x^2+b(t)x+c(t),
\label{eq09}
\end{equation}
where the function $a=a(t)$ can be expressed in terms of $r$ and $f$ using the formula $a=-r'_t/(2rf)$.
Substituting \eqref{eq09} into the first equation \eqref{eq04}, we arrive at a quadratic equation with respect to $x$, the coefficients of which depend on time.
As a result, the first equation of the system is reduced to a quadratic equation with respect to $x$, the coefficients of which depend on time. By equating the functional coefficients of this quadratic equation to zero and adding the second equation of the system, which in this case depends only on $t$, we obtain the following system consisting of four ODEs:
\begin{equation}
\begin{aligned}
a'_t&=-4a^2f(r),\\
b'_t&=-4abf(r),\\
c'_t&=-b^2f(r)+g(r),\\
r'_t&=-2arf(r).
\end{aligned}
\label{eq10}
\end{equation}
Here the three first equations were divided by $r$.

From the first, second and fourth equations of system \eqref{eq10} we can obtain two integrals
\begin{equation}
a=C_1r^2,\quad \ b=C_2r^2,
\label{eq10*}
\end{equation}
where $C_1$ and $C_2$ \arbs.
Substitute the first expression \eqref{eq10*} into the first equation \eqref{eq10}.
Integrating the resulting ODE, we find the dependence $r=r(t)$ in implicit form
\begin{equation}
\int\frac{dr}{r^3f(r)}=-2C_1t-C_3,
\label{eq11}
\end{equation}
where  $C_3$ \arb.
Integrating the third equation of system \eqref{eq10}, we obtain
\begin{equation}
c=-C_2^2\int r^4f(r)\,dt+\int g(r)\,dt+C_4.
\label{eq12}
\end{equation}
For simplicity, assuming in \eqref{eq10*} and \eqref{eq12} $C_2=0$ and substituting the obtained expressions into \eqref{eq09},
we find the functions that determine solution \eqref{eq02}:
\eqref{eq02}:
\begin{equation}
r=r(t),\quad \ \varphi=C_1r^2(x+A)^2+\int g(r)\,dt+C_4,
\label{eq12*}
\end{equation}
where the function $r=r(t)$ is given implicitly by expression \eqref{eq11}.
Note that an arbitrary constant $A$ has been added to the second formula,
since the system \eqref{eq04} is invariant with respect to an arbitrary constant shift along the spatial variable.
\medskip

\textit{Example 6.}
For equation \eqref{eq01} with the power dispersion $f(r)=ar^k$ in solution \eqref{eq12*} we must set
\begin{align*}
r(t)=\bl[a(k+2)(2C_1t+C_3)\br]^{-\tfrac 1{k+2}}.
\end{align*}
This formula is derived from \eqref{eq11}.
\goodbreak

\section{Solutions that are nonlinear superpositions of traveling waves}\label{s:4}

The system of PDEs \eqref{eq04} admits an exact solution of the form
\begin{equation}
r=r(z),\quad \ \varphi=C_1t+C_2x+\theta(z),\quad \ z=x-\lambda t,
\label{eq05*}
\end{equation}
where $C_1$, $C_2$, and $\lambda$ \arbs.
The special case $C_1=C_2=0$ in \eqref{eq05*} corresponds to a traveling wave solution.

Substituting \eqref{eq05*} into \eqref{eq04}, we obtain the nonlinear system of two ODEs:
\begin{equation}
\begin{aligned}
-r(C_1-\lambda\theta'_z)+h_{zz}''-h(C_2+\theta'_z)^2+rg(r)&=0,\\
-\lambda r_z'+2h_z'(C_2+\theta'_z)+h\theta_{zz}''&=0,\quad \ h=rf(r).
\end{aligned}
\label{eq09*}
\end{equation}

Note that the substitution $\xi=\theta'_z$ allows us to reduce the order of the ODE system \eqref{eq09*} by one.

It is easy to verify that the second equation \eqref{eq09*} admits the first integral
\begin{equation}
-\lambda\int h(r)\,dr+C_2h^2+h^2\theta'_z=C_3,\quad \ h=rf(r),
\label{eq20}
\end{equation}
where $C_3$ \arb.
Eliminating further the derivative $\theta'_z$ from the first equation \eqref{eq09*} using \eqref{eq20}, we arrive at a second-order nonlinear autonomous ODE (that does not explicitly depend on $z$) of the form
\begin{equation}
h_{zz}''=\Phi(h)+\Psi(r),
\label{eqq40}
\end{equation}
where the following notations are used:
\begin{equation}
\begin{aligned}
\Phi(h)&=C_3^2h^{-3},\quad h=rf(r),\\
\Psi(r)&=(C_1+\lambda C_2)r-C_3\lambda rh^{-2}-\lambda^2rh^{-2}H+2C_3\lambda h^{-3}H\\
&\quad\, +\lambda^2h^{-3}H^2-rg(r),\quad \ H=\int h\,dr.
\end{aligned}
\label{eqq41}
\end{equation}
The appendix shows that the general solution of equation \eqref{eqq40}--\eqref{eqq41} can be expressed in quadratures and presented in an implicit form.

By resolving relation \eqref{eq20} with respect to $\theta'_z$, and then integrating, we can find the function $\theta=\theta(z)$,
$$
\theta=C_3\int h^{-2}dz+\lambda\int h^{-2}H\,dz-C_2z+C_4,
$$
where $C_4$ \arb.
\medskip

\textit{Remark 4.}
For the physical interpretation of solution \eqref{eq05*}, it is convenient to represent the phase $\varphi$ in an equivalent form in two ways:
\begin{align*}
\varphi&=C_2(x-\lambda_2 t)+\theta(x-\lambda_1t),\qquad \lambda_1=\lambda,\quad \lambda_2=-C_1/C_2;\\
\varphi&=(C_2+C_1\lambda^{-1})x+\theta_1(x-\lambda t),\quad \ \theta_1(z)=\theta(z)-C_1\lambda^{-1}z.
\end{align*}
In the first case solution \eqref{eq05*} can be interpreted as a nonlinear superposition of two traveling waves with velocities $\lambda_1$ and $\lambda_2$, and in the second case as a nonlinear superposition of a stationary wave and a traveling wave with velocity $\lambda$.
\medskip

\textit{Remark 5.}
Solution \eqref{eq05*} can be obtained, for example, using the following reasoning.
First of all one can note,
that one of the two arbitrary constants $C_1$ or $C_2$ in \eqref{eq05*} can be taken equal to zero without loss of generality (this fact corresponds to the redefinition of the function $\theta$).
Next, we pass in equation \eqref{eq01} or system \eqref{eq04} from the variables $x$, $t$ to new independent variables $z$, $t$, where $z=x-\lambda t$.
Then, by analogy with \eqref{eq02ab}, we introduce the additional relation $|u|=p(z)$).
After this, the procedure of separation of variables is used.

We will also describe another simpler way to obtain solution \eqref{eq05*}.
First, in Eq. \eqref{eq01} we make the substitution $u=e^{i(C_1t+C_2x)}w$, where $C_1$ and $C_2$ are arbitrary real constants.
Then a traveling wave solution of the transformed equation is sought, i.e. it is assumed that $w=W(z)$, where $z=x-\lambda t$.

\section{Brief conclusion}

For the first time, the
general nonlinear Schr\"odinger equation has been investigated, the dispersion and potential of which are specified by two arbitrary functions.
New several closed-form solutions of this PDE have been found, which are expressed in quadratures or elementary functions.
One-dimensional reductions have been described, which leads the considered nonlinear PDE to simpler ODEs.
The obtained solutions can be used as test problems for numerical methods of integrating nonlinear PDEs of mathematical physics.
It is important to note that these exact solutions are valid for two arbitrary functions $f(z)$ and $g(z)$ included in the general nonlinear Schr\"odinger equation \eqref{eq01}, so they can be used for a wide variety of problems, specifying a specific form of these functions.

\section*{Data availability}

No data was used for the research described in the article.

\section*{Declaration of competing interest}

The authors declare that there is no conflict of interest.

\section*{Acknowledgements}

The study was supported by the Ministry of Education and Science of the Russian Federation
within the framework of the state assignments No. 124012500440-9 and No. FSWU-2023-0031.

\section*{Appendix. Solutions of special second-order ODEs}

Let us consider a class of second-order autonomous ODEs of the special form
\begin{equation}
h_{xx}''=\Phi(h)+\Psi(r),\quad \ h=\Theta(r),
\label{eqq08*}
\end{equation}
where $\Phi(h)$, $\Psi(r)$, and $\Theta(r)$ are  given functions.
ODE \eqref{eq08} is a particular case of equation \eqref{eqq08*} with
\begin{equation}
\Phi(h)=C_2^2h^{-3},\quad \ \Psi(r)=C_1r-rg(r),\quad \ \Theta(r)=rf(r),
\label{eqq08**}
\end{equation}

For $\Psi(r)\equiv 0$, the general solution of the ODE \eqref{eqq08*} can be represented in implicit form \cite{polzai2018}:
\begin{equation}
\int\BL[2\int\Phi(h)\,dh+A_1\BR]^{-1/2}dh=A_2\pm x,\quad \ h=\Theta(r),
\label{eq08*a}
\end{equation}
where  $A_1$ and $A_2$ \arbs.
In the general case, in the second term of the right-hand side of equation \eqref{eqq08*} we pass from the variable $r$ to the variable $h$, setting $\Psi_1(h)=\Psi(r)$, where $h=\Theta(r)$. We have
\begin{equation}
h_{xx}''=\Phi(h)+\Psi_1(h).
\label{eq08*b}
\end{equation}
The general solution of this equation is determined by the formula \eqref{eq08*a}, in which $\Phi(h)$ must be replaced by $\Phi(h)+\Psi_1(h)$, or
\begin{equation}
\int\BL[2\int\Phi(h)\,dh+2\int\Psi_1(h)\,dh+A_1\BR]^{-1/2}dh=A_2\pm x,\quad \ h=\Theta(r),
\label{eq08*c}
\end{equation}
We do not change the first inner integral in \eqref{eq08*c}, and in the second inner integral we return from the variable $h$ to the variable $r$.
Taking into account the relations $\Psi_1(h)=\Psi(r)$ and $dh=\Theta'_r(r)\,dr$, we finally obtain
\begin{equation}
\int\BL[2\int\Phi(h)\,dh+2\int\Psi(r)\Theta'_r(r)\,dr+A_1\BR]^{-1/2}dh=A_2\pm x,\quad \ h=\Theta(r).
\label{eq08*d}
\end{equation}

Substituting expressions \eqref{eqq08**} into \eqref{eq08*d}, we find the general solution of Eq.~\eqref{eq08}:
\begin{equation*}
\begin{gathered}
\int\BL[-C_2^2h^{-2}+2\int[C_1r- rg(r)]\,dh+A_1\BR]^{-1/2}\,dh=A_2\pm x,\\
h=rf(r),\quad \ dh=[f(r)+rf'_r(r)]\,dr.
\end{gathered}
\end{equation*}

\renewcommand{\refname}{References}

\end{document}